\documentclass[pre,amsmath,twocolumn,showpacs,eqsecnum,preprintnumbers,twoside]{revtex4}
\usepackage{graphicx}
\begin{document}

\title{One-Dimensional Kronig-Penney Model with Positional
       Disorder:\\ Theory versus Experiment}

\author{G.~A.~Luna-Acosta, F.~M.~Izrailev}
\affiliation{Instituto de F\'{\i}sica, Universidad Aut\'{o}noma de
            Puebla, Apartado Postal J-48, Puebla, Pue., 72570, M\'{e}xico}
\author{N.~M.~Makarov}
\affiliation{Instituto de Ciencias, Universidad Aut\'{o}noma
            de Puebla,\\ Priv. 17 Norte No 3417, Col. San Miguel
            Hueyotlipan, Puebla, Pue., 72050, M\'{e}xico}
\author{U.~Kuhl, H.-J. St\"{o}ckmann}
\affiliation{Fachbereich Physik der Philipps-Universit\"{a}t Marburg, Renthof 5, D-35032, Germany.}
\date{\today}

\begin{abstract}
We study the effects of random positional disorder in the
transmission of waves in a 1D Kronig-Penny model. For weak disorder
we derive an analytical expression for the localization length and relate it to the transmission coefficient for finite samples. The obtained results describe very well the experimental frequency dependence of the transmission in a microwave realization of the model. Our results can be applied both to photonic crystals and semiconductor super lattices.
\end{abstract}


\maketitle

\section{Introduction}
In recent years there is a high activity in the study of wave (electron) propagation through one-dimensional periodic structures (see, for example, Ref.\cite{mar08} and references therein). Much is already known about band structures of perfectly propagating waves in strictly periodic and relatively simple devices, and one of the current interests is the influence of random imperfections that are commonly present in real experiments. These imperfections are originated, for example, from the variations of the medium parameters
such as the dielectric constant, magnetic permeability, barrier widths or
heights \cite{mcg93,smi00,she01,par03,esm06,nau07,pon07,asa07}.

The analysis of scattering properties of periodic-on-average (when periodic systems are slightly affected by a disorder) models with various kinds of disorder is mainly related to numerical methods. It is obvious that giving important results for specific models and parameters, the numerical approaches can not serve as a guide for the  understanding of generic properties caused by disorder. In this paper we try to fill this gap in the theory by the derivation of the localization length for the 1D Kronig-Penney model, relating it to the properties of transmission through a finite number of disordered barriers.

Our analytical results are compared with the experimental data obtained for a single-mode microwave guide. We show that in spite of the standard restrictions of analytical results (restricted to infinite samples and weak disorder), comparison between theory and experiment is quite good. This fact is highly non-trivial since the experimental data are strongly influenced by absorbtion in the waveguide walls; effect that is also not taken into account analytically.

Our study is relevant to other types of 1D stratified media, for example, to electron transport through  random superlattices \cite{don98b} (disordered arrays of semiconductor quantum wells/barriers) or acoustic waves in random layered media \cite{bal85}. Also, similar properties of the transmission are expected to occur in the 1D {\it quantum} Kronig-Penney model (with alternating rectangular wells and barriers).

In Sect. II the model is specified and the transfer matrix equations are derived. In Sect. III the experimental setup is briefly discussed and the numerical simulations for the transmission coefficient are compared with experimental results for the case of an array of 26 cells and different amounts of positional disorder. In Sect. IV we present the main experimental results and discuss some of the properties of transmission. In Sect. V we derive, for the regime of weak disorder, the analytical expression for the logarithm of the transmission in connection with the inverse localization length. We compare there too the numerical simulations and analytical results with the experimental data and show the effectiveness of our analytical approach. In Sect. VI, we summarize our results.

\section{Model}

\begin{figure}
\includegraphics[width=\columnwidth]{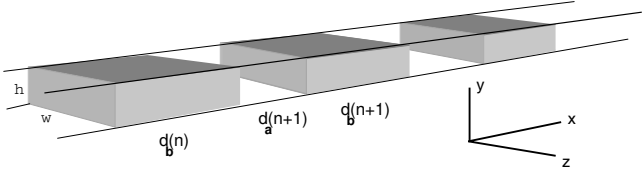}
\caption{\label{fig:fig1} (color online) Kronig-Penney
model with teflon bars of constant length $d_b(n)=d_b(n+1)=d_b$, and air spacings $d_a(n)$ defined by Eq.~(\ref{dan-db}).}
\end{figure}

We consider an array formed by two alternating dielectric slabs with refractive indices $n_a$ and $n_b$ placed in an electromagnetic metallic-wall waveguide of constant width $w$ and height $h$, see Fig.~\ref{fig:fig1}. For convenience, the layers with the refractive index $n_a$ ($n_b$) shall be referred as to the $a$-layers ($b$-layers). The lengths of the $n$th $a$- and $b$-layers are denoted, respectively, by $d_{a}(n)$ and $d_b(n)$. The positional disorder in our model consists in randomly varying lengths of \emph{only} one type of layer, say the $a$-layer, such that
\begin{equation}\label{dan-db}
d_a(n)=d_a+\sigma\eta(n),\ \langle d_a(n)\rangle=d_a,
\quad d_b(n)=d_b.
\end{equation}
Here $\sigma$ is r.m.s deviation of $d_a(n)$ and $\sigma^2$ its variance. Hence, $\eta(n)$ is a sequence with zero average and unit variance. In this work we assume that $\eta(n)$ is random uncorrelated , i.\,e.
\begin{equation}\label{corr-sym}
\langle\eta(n)\eta(n')\rangle=\delta_{nn'},\qquad \langle\eta(n)\rangle=0.
\end{equation}
The angular brackets $\langle\ldots\rangle$ stand for a statistical
average over different realizations of randomly layered
structure. Note that the random structure is \emph{periodic on
average} with the period $d=d_a+d_b$.

In this work we shall treat the lowest TE mode of frequency $\nu$ whose
electric field $\mathbf{E}$ is defined by
\begin{equation}\label{E-def}
E_y=\sin(\pi z/w)\Psi(x),\quad E_x=E_z=0,
\end{equation}
(see Fig.~\ref{fig:fig1}). Within every $a$- or $b$-layer, the function $\Psi(x)$ obeys the 1D Helmholtz equation,
\begin{equation}\label{SchrEq-ab}
\left(\frac{d^2}{dx^2}+k_{a,b}^2\right)\Psi_{a,b}(x)=0,
\end{equation}
with the wave numbers
\begin{equation}\label{kab-def}
k_{a,b}=\frac{2\pi}{c}\sqrt{n_{a,b}^2 \nu^2-(c/2w)^2}.
\end{equation}
In the incoming-outgoing wave representation, the solution
of Eq.~\eqref{SchrEq-ab} for the $n$th elementary cell reads
\begin{eqnarray}\label{PsiAB-exp1}
\Psi_{a_n}(x)&=&A^{+}_{n}\exp\left\{ik_a[x-x_a(n)]\right\}
\nonumber\\[6pt]
&&+A^{-}_{n}\exp\left\{-ik_a[x-x_a(n)]\right\}
\label{PsiAn-exp}
\end{eqnarray}
inside the $a_n$-layer, $x_a(n)\leq x\leq
x_b(n)$, and
\begin{eqnarray}\label{PsiAB-exp2}
\Psi_{b_n}(x)&=&B^{+}_{n}\exp\left\{ik_b[x-x_b(n)]\right\}
\nonumber\\[6pt]
&&+B^{-}_{n}\exp\left\{-ik_b[x-x_b(n)]\right\}
\label{PsiBn-exp}
\end{eqnarray}
inside the $b_n$-layer, $x_b(n)\leq x\leq
x_a(n+1)$.
Here $A^{\pm}_{n}$ and $B^{\pm}_{n}$ are complex amplitudes of the forward/backward traveling wave, the coordinates $x_a(n)$, $x_b(n)$ denote the left-hand boundaries of the $a_n$ and $b_n$ layers, respectively.

With the use of the continuity conditions for the wave function $\Psi_{a,b}(x)$ and its derivative at the boundaries $x=x_b(n)$ and
$x=x_a(n+1)$ one can obtain the transfer relation for the
amplitudes $A^{\pm}_{n+1}$ and $A^{\pm}_n$ of two adjacent cells,
\begin{equation}\label{An+1An-exp}
\left(\begin{array}{c}A^{+}_{n+1}\\A^{-}_{n+1}\end{array}\right)=
\left(\begin{array}{cc}Q_{11}(n)&Q_{12}(n)\\[6pt]
Q_{21}(n)&Q_{22}(n)\end{array}\right)
\left(\begin{array}{c}A^{+}_n\\A^{-}_n\end{array}\right).
\end{equation}
The transfer matrix $\hat{Q}(n)$ has the following elements,
\begin{subequations}\label{Q-def}
\begin{eqnarray}
Q_{11}(n)&=&[\cos(k_bd_b)+i\alpha_+\sin(k_bd_b)]\exp[ik_ad_a(n)]
\nonumber\\[6pt]
&=&Q^*_{22}(n);\label{Q11}\\[6pt]
Q^*_{12}(n)&=&i\alpha_-\sin(k_bd_b)\exp[ik_ad_a(n)]
\nonumber\\[6pt]
&=&Q_{21}(n).\label{Q12}
\end{eqnarray}
\end{subequations}
Here the asterisk stands for the complex conjugation, and we
introduced the parameters $\alpha_\pm$,
\begin{equation}\label{alpha-def}
\alpha_\pm=\frac{1}{2}\left(\frac{k_a}{k_b}\pm\frac{k_b}{k_a}\right),\qquad
\alpha_+^2-\alpha_-^2=1.
\end{equation}
The determinant of $\hat{Q}(n)$ is equal to unit, $\det{\hat{Q}(n)}=1$. Note that the transfer matrix $\hat Q(n)$ differs from cell to cell only in the phase factor $\exp[ik_ad_a(n)]$.

The transfer matrix equation for the array of $N$ cells with or without positional disorder is
\begin{equation}\label{AN+1AN-exp}
\left(\begin{array}{c}A^{+}_{N+1}\\A^{-}_{N+1}\end{array}\right)=\hat Q^N \left(\begin{array}{c}A^{+}_1\\A^{-}_1\end{array}\right),
\end{equation}
where,
\begin{equation}\label{QN-def}
\hat Q^N=\hat Q(N)\hat Q(N-1)..\hat Q(n)...\hat Q(2)\hat Q(1).
\end{equation}
All matrices $\hat{Q}(n)$ ($n=1,2,\dots,N$) have the same form \eqref{Q-def}, only differing in the values of $d_a(n)$. In our following numerical simulations and experimental set-up we have $A^{-}_{N+1}=0$. Thus the transmittance of $N$ cells is given by
\begin{equation}\label{TN-gen}
T_N\equiv|A^{+}_{N+1}/A^{+}_1|^2=|Q^N_{11}|^{-2}=|S_{12}^N|^2,
\end{equation}
where $S_{12}^N$ is the scattering matrix element in the relation
\begin{equation}\label{AN+1An-Scat}
\left(\begin{array}{c}A^{-}_1\\A^{+}_{N+1}\end{array}\right)=
\left(\begin{array}{cc}S_{11}^N&S_{12}^N\\[6pt]
S_{21}^N&S_{22}^N\end{array}\right)
\left(\begin{array}{c}A^{+}_1\\A^{-}_{N+1}\end{array}\right).
\end{equation}

In the case of \emph{no disorder}, $\eta(n)=0$ , the length of $a$-layer does not depend on the cell number $n$, $d_a(n)=d_a$. Therefore, the unperturbed transfer matrix $\hat{Q}^{(0)}$ is described by Eq.~\eqref{Q-def} with $d_a(n)$ replaced by the constant length $d_a$.
As is known (see, e.g., Ref.~\cite{mar08}), the transmission through $N$
\emph{identical} cells is expressed in closed form as
\begin{equation}\label{TN-0}
T^{(0)}_N=\frac{1}{1+ \left|Q^{(0)}_{12} \frac{\sin(N\kappa
d)}{\sin(\kappa d)}\right|^2}=|S^{(0)N}_{12}|^2,
\end{equation}
where $\kappa$ is the Bloch wave number defined by the dispersion relation
\begin{equation}\label{DispRel}
\cos(\kappa d)=\cos(k_ad_a)\cos(k_bd_b)
-\alpha_+\sin(k_ad_a)\sin(k_bd_b).
\end{equation}

Expression~\eqref{TN-0} indicates that the transmission is perfect ($ S^{(0)N}_{12}=1$) for all $N$ where $Q^{(0)}_{12}=0$ or when $\sin(N\kappa d)/\sin(\kappa d)=0$. The former occurs at the
Fabry-Perot resonances $k_bd_b=m\pi$ in the $b$-layers. The latter produces $N-1$ Fabry-Perot oscillations in each spectral band, associated with the total system length $Nd$. We shall refer to the resonances $k_bd_b=m\pi$ as ``teflon resonances'' since in the experiment the $b$-slabs are made of teflon, whereas the $a$-slabs are just air.
\vspace{1.3cm}

\begin{figure}
\mbox{\includegraphics[height=0.55\columnwidth]{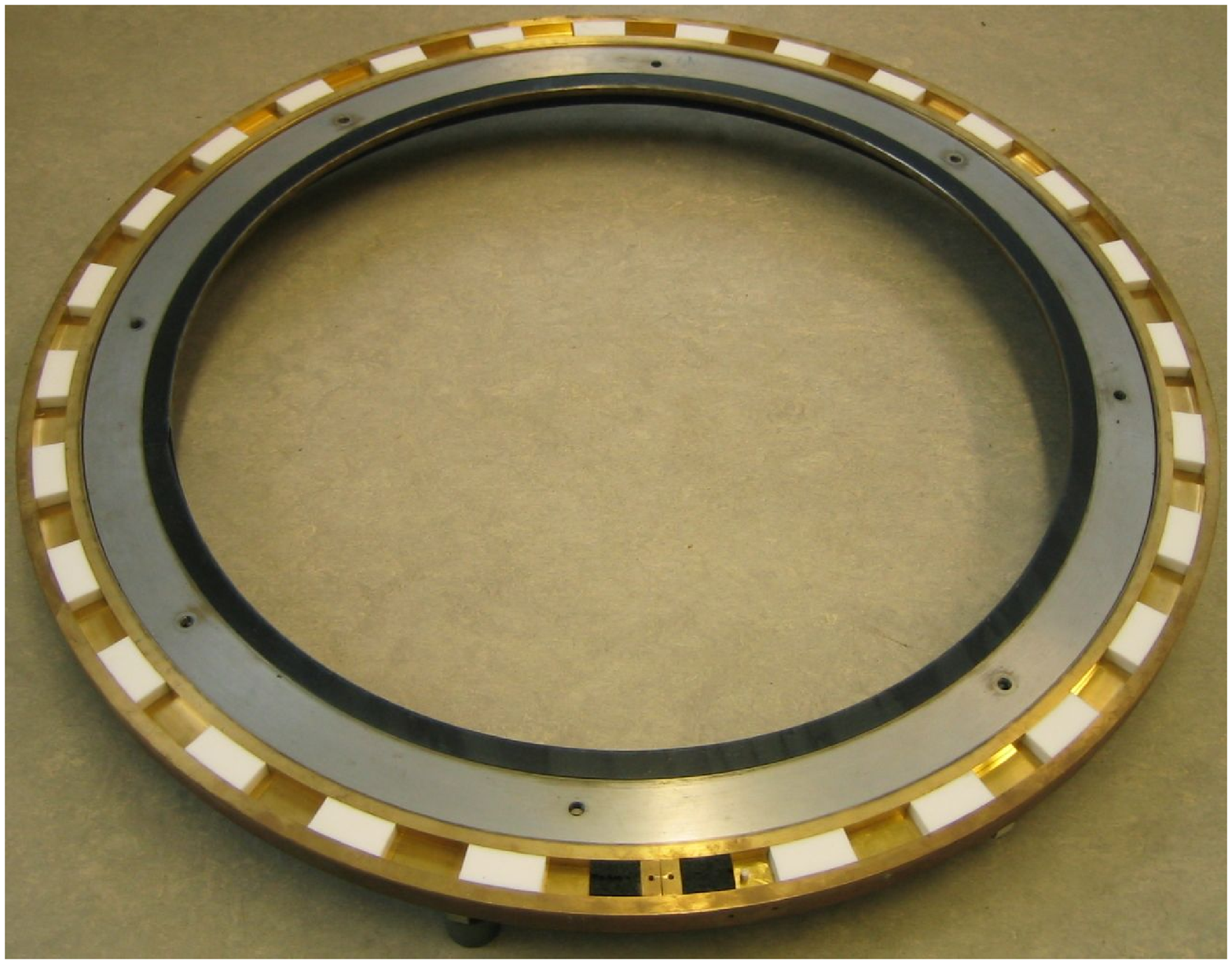}
\includegraphics[height=0.55\columnwidth]{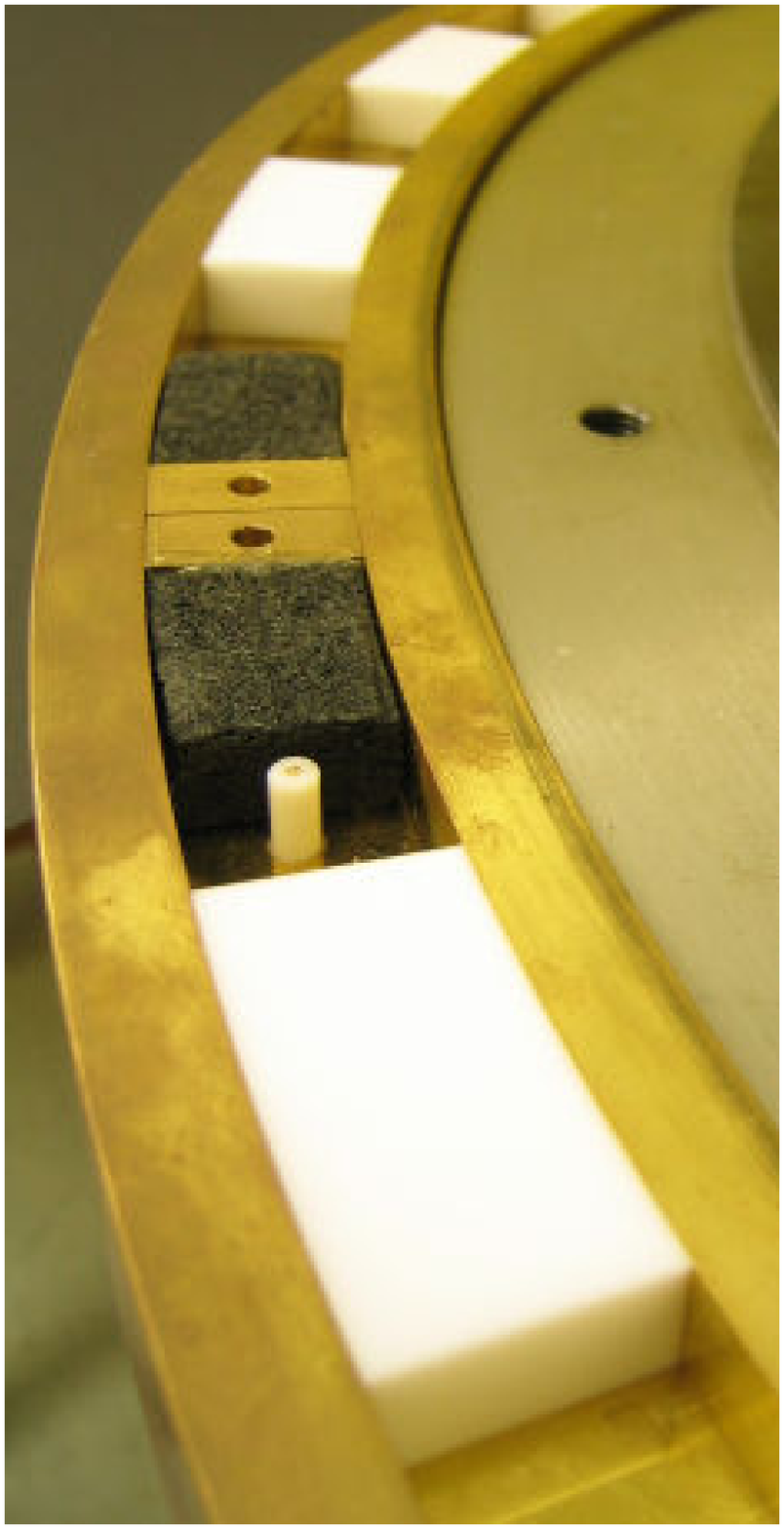}}
\caption{\label{fig:fig2}(color online) The microwave guide. Left: overview of the brass waveguide of perimeter $P=234$ cm, with height $h=1$ cm and width $w=2$ cm. Right: an enlargement of the waveguide showing the two carbon
pieces (black) and one of the electric dipole antennas. The other lies on a brass lid (not shown) inserted near the other carbon
piece. The white pieces are the teflon bars.}
\end{figure}

\section{Experimental Setup}

In Fig.~\ref{fig:fig2} we show the experimental setup of a microwave ring guide
of height $h=1$ cm, width $w=2$ cm and perimeter $P=234$ cm. The waveguide consists of $N=26$ cells, where the $b-$layers are pieces of teflon of length $d_b=4.078$ cm and the refractive index $n_b=\sqrt{2.08}$.
Two electric dipole antennas connect to a network analyzer. Also
shown are the two carbon pieces used to absorb the electric field
at both ends of the waveguide and thus mimic two infinite leads
connected to each side of the array of teflon pieces and air
segments. The frequency range is 7.5 to 15\,GHz,
corresponding to wave lengths from 4 to 2\,cm. This arrangement
has been used to study the transport effects of single impurities in the photonic Kronig-Penny model \cite{lun08}. Earlier, in an analogous model with metallic screws instead of teflon pieces, the microwave realization of the Hoftstadter butterfly \cite{kuh98b} was studied. The same configuration (metallic screws) was used to investigate transport properties of on-site correlated disorder \cite{kuh00a,kro02,kuh08a}.

Obviously this waveguide is not rectilinear; in the experiment the teflon
pieces and air segments are not perfect parallelepipeds: one side is longer than the other by 5 percent. However since the
perimeter (234\,cm) is much larger than the wave length of the
electric field even in the regime of the first mode, it is expected
that the rectilinear model is a good approximation. In fact,
as shown in detail in \cite{lun08}, a good quantitative agreement is found by defining an effective length of the teflon pieces and the
air segments that are $1.95$ percent larger than the smaller side of the teflon pieces and air segments. For example, the length of the
smaller (inner) length of the teflon pieces used in the experiments reported here is 4 cm, so the effective width $d_a$ we use in our
calculations is 4.078\,cm. We remark that this value is found by
best fitting and is the only fitting, good for the whole frequency range of all our results presented here.

\begin{figure}[h!]
\includegraphics[width=\columnwidth]{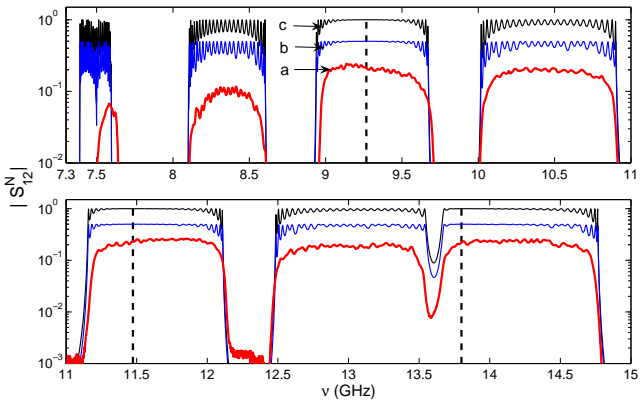}
\caption{\label{fig:fig3} (color online) Transmission spectrum for a $26$-cells array. Top: $7.3<\nu<11$ GHz. Bottom: $11<\nu<15$ GHz. Curve (a): experimental data for the periodic array.  Curve (b): $|S_{12}^N|$, multiplied by $1/2$, for a slightly random ($\epsilon=0.049$) array. Curve (c): $|S_{12}^{(0)N}|$ for the perfectly periodic array ($d_a=d_b=4.078$ cm).  Dashed vertical  lines mark the position of the teflon resonances.}
\end{figure}

In Fig.~\ref{fig:fig3}, curve (a), we plot the experimentally measured value of $|S_{12}^N|$ (in what follows, the {\it transmission spectrum}) for the 26-cells \emph{periodic array}. Curve (c) is the transmission spectrum $|S^{(0)N}_{12}|$, calculated  according to Eqs~\eqref{TN-0}, \eqref{DispRel} and unperturbed Eq.~\eqref{Q-def}. Note that the experimental transmission spectrum is about $1/5$ of the theoretical one. This decrease of the signal is due to absorbtion by the metal walls of the waveguide. The value of the transmission spectrum is roughly constant over the frequency range, and can be taken into account phenomenologically by introducing an absorbtion factor. However, it is not the purpose of our work here to study the absorbtion effects. Our question is the global frequency dependence of the transmission spectrum on the frequency, giving us a possibility to reveal resonance effects and the role of disorder. Note that the band structure of the spectrum remains practically the same in spite of a strong absorbtion (see discussion in Ref.~\cite{kuh00a,kro02,kuh08a}).

Inspection of the transmission spectrum $|S^{(0)N}_{12}|$ of the perfectly periodic array, Fig.~\ref{fig:fig3}, curve (c), demonstrates the effect of the teflon resonances on the transmission bands. We see two types of bands; namely, bands 1,2,4, and 6 show the $N-1=25$ oscillations mentioned above, whereas bands 3, 5 and 7 are flat ($|S^{(0)N}_{12}|\approx 1$) around the resonance since $|Q^{(0)}_{12}|$ is zero at the resonance, and very small in some neighborhood around it.  For future reference we shall refer to the second type of  bands as the "resonance bands". Clearly, these bands disappear when $d_b\to0$, turning into those with $N-1$ oscillations occurring for delta-function potentials.

In accordance with Eq.~\eqref{kab-def}, the first (lowest) mode in the \emph{air spacings} ($a$-slabs) opens at the cut-off frequency $\nu_a^{cut}=(c/2wn_a)=7.5$ GHz while in the \emph{teflon $b$-layers} it opens at $\nu_b^{cut}=(c/2w n_b)=5.2$ GHz, which is less than $\nu_a^{cut}$ ($n_a<n_b$). On the other hand,
Fig.~\ref{fig:fig3} specifies the bottom of the first transmission band at $\nu_1^{bot}=7.387$ GHz. So that the real cutoff $\nu_1^{bot}$ pertains to the interval $\nu_b^{cut}<\nu_1^{bot}<\nu_a^{cut}$ where the wave number $k_a$ in the air spacings is purely imaginary, $k_{a}=i(2\pi/c)\sqrt{(c/2w)^2-\nu^2}$. The transmission in this regime is due to tunneling through the air spacings. This fact makes the profile of the first band (see Fig.~\ref{fig:fig3}, curve (c)) somewhat special: there is a dip in the transmission band right at the frequency where the first mode opens in the air.

Note that except for the first band, the positions and widths of the transmission gaps as well as the band profiles of the experimental curve, are well reproduced by the theoretical calculations. The discrepancy for the first band is understood since for low frequencies the wave length is not sufficiently small compared to the perimeter of the circular waveguide, and hence the rectilinear waveguide model fails. For the second and higher bands, the agreement is better and the experimental curve does show some evidence of the small $N-1$ band oscillations predicted by the model. Clearly, these do not appear as perfectly regular oscillations and this irregularity may be caused by experimental imperfections due to variations  in the length and positions of the teflon pieces.

The maximum deviation $|d_b(n)-d_b|_{max}$ in the length of the teflon pieces  is about 0.01\,cm  and the maximum deviation $|d_a(n)-d_a|_{max}$ in the air spacings, due to the placement of the teflon pieces, is estimated to be about $0.04$ cm. Are imprecisions of this order sufficient to break the regularity of the oscillations?

To check this, we performed a  simulation with a disordered sequence assuming $|d_a(n)-d_a|_{max}=0.04$\,cm and an error described by a random sequence $\eta(n)$ with a uniform random distribution in accordance with Eqs.~\eqref{dan-db}, \eqref{corr-sym},
\eqref{QN-def} and \eqref{TN-gen}. The result is plotted in Fig.~\ref{fig:fig3}, curve (b). This curve has been multiplied by $0.5$ in order to show it together with the perfectly regular case and the experimental data. Inspection shows that indeed the assumed small error is enough to break the regularity of the band oscillations giving a better agreement with the experimental data of the supposedly regular array.

\section{Disordered Array}

Let us now move to intentionally disordered arrays, with the lengths of all teflon bars
constant, $d_b(n)=d_b$ while the air layers have random lengths given by Eqs.~\eqref{dan-db}, \eqref{corr-sym}. In  our experimental and
numerical calculations, the sequence $\eta(n)$ is an uncorrelated  random
function uniformly distributed in the interval $[-\sqrt 3,\sqrt 3]$, with unit variance.

{\it Apriori} it is not known how large a random deviation from the average value $d_a$ should be to observe weak, medium or strong disorder effects in the transmission. We tentatively classify the amount of disorder by the value of the maximum deviation from the average length of the air spacing divided by the average length of the cells,
\begin{equation}\label{DisMeasure}
\epsilon\equiv\frac{|d_a(n)-d_a|_{max}}{d}=\sigma\sqrt{3}/d.
\end{equation}

Table~\ref{tab:Parameter} shows the values of $\epsilon$ we consider in this work, together with corresponding values of relative r.m.s. $\sigma/d$ and $(\sigma/d)^2$. The latter quantity is needed to ease the comparison with analytical results obtained below. The case of $\epsilon=0.49\times10^{-2}$ was discussed above to simulate the errors in the experimental setup. We call it the case of extremely weak disorder. Similarly, $\epsilon=3.0\times10^{-2}$, $12.3\times10^{-2}$ and $49.0\times10^{-2}$ cm, respectively, are called the weak, medium, and strong disordered cases.

\begin{table}
\begin{tabular}{|c|r|r|c|c|}\hline
 case   & $\epsilon/10^{-2}$ & $\frac{\sigma}{d}\left(=\frac{\epsilon}{\sqrt{3}}\right)$/cms &
$\left(\frac{\sigma}{d}\right)^2$\\ \hline
very weak &  0.49 &  $0.28 \cdot 10^{-2}$ & $8.0 \cdot 10^{-6}$ \\ \hline
weak      &  3.00 &  $1.77 \cdot 10^{-2}$ & $3.0 \cdot 10^{-4}$ \\ \hline
medium    & 12.30 &  $7.07 \cdot 10^{-2}$ & $5.0 \cdot 10^{-3}$\\ \hline
strong    & 49.00 & $28.30 \cdot 10^{-2}$ & $8.0 \cdot 10^{-2}$\\ \hline
\end{tabular}
\caption{\label{tab:Parameter} Parameter values of random disorder. Here $d=d_a+d_b=2d_a\approx 8.16$\,cm}
\end{table}

\begin{figure*}
\includegraphics[width=\textwidth]{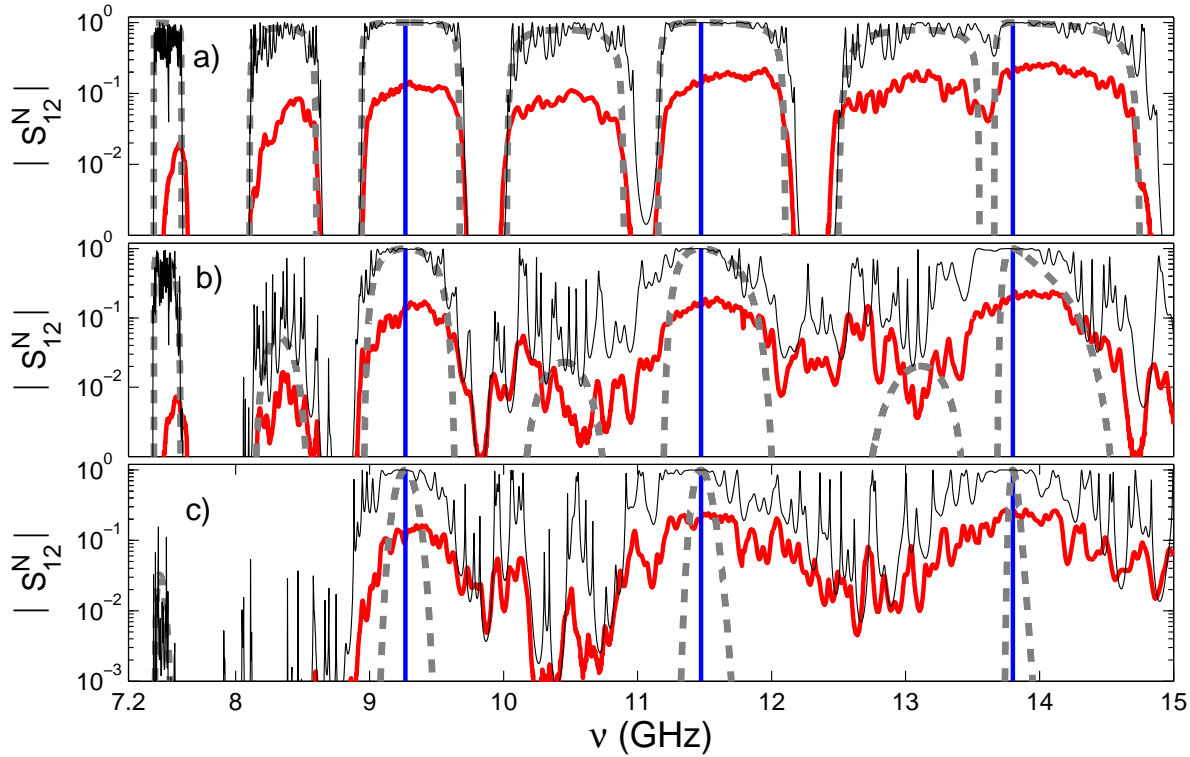}
\caption{\label{fig:fig4}(color online) Transmission spectrum $|S_{12}^N|$ for $26$-cells arrays: (a) weak ($\epsilon=3.0\times10^{-2}$), (b) medium ($\epsilon=12.3\times10^{-2}$), and (c) strong disorder ($\epsilon=49.0\times10^{-2}$). Transfer matrix calculations are shown by thin black solid curves, and experimental measurements by thick red solid curves. Dashed curves correspond to the analytical expression (\ref{ave-S12}) for the inverse localization length. Perpendicular lines mark the position of the teflon resonances.}
\end{figure*}

Fig.~\ref{fig:fig4} shows the transmission for the array of $26$-cells
with the positional disorder. Compared with the case of weak disorder (Fig.~\ref{fig:fig4}a), for medium disorder (Fig.~\ref{fig:fig4}b) only the first two gaps are clearly distinguishable; the third only partially. There is no trace of the $N-1$ oscillations in the transmission bands, and the second, fourth, and sixth  transmission bands have decayed substantially. However, remnants of the \emph{resonance bands} are still recognized, and so this can be considered the regime of medium disorder.
For strong disorder (Fig.~\ref{fig:fig4}c) the first two transmission bands have disappeared. There is no longer any evidence of the band structure of the unperturbed array. But still the transmission spectrum is close to one in the vicinity of the teflon resonances.

\begin{figure*}
\includegraphics[width=\textwidth]{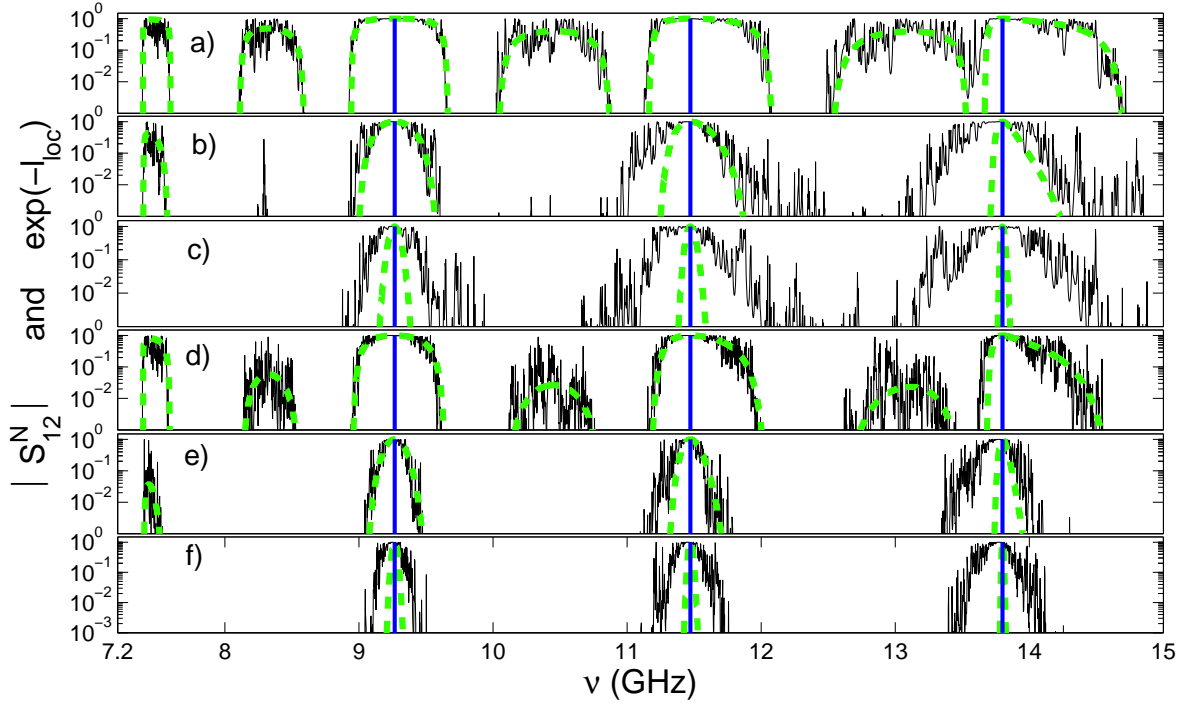}
\caption{\label{fig:fig5}(color online) Transmission spectrum $|S_{12}^N|$ (solid curves) and $\exp(-l_{loc})$ (dashed curves) for $N=100$ and $N=400$: (a) to (c)
N=$100$ cells; a) weak, b) medium and c) strong disorder. d) to f)
N=$400$ cells; d) weak, e) medium and f) strong disorder.}
\end{figure*}

The results so far discussed pertain to the array of $N=26$ cells so the question arises about the effects on larger arrays. Our experimental setup does not allow for the implementation of much larger arrays. However, given that the numerical simulations are in good correspondence with the experimental data for $N=26$ cells, we now consider only numerically, and later analytically, larger arrays for the same three cases: weak, medium, and strong disorder.
In Figs.~\ref{fig:fig5}a -- \ref{fig:fig5}c we plot  $|S_{12}^N|$ for an array with $N=100$ cells, for weak (Fig.~\ref{fig:fig5}a), medium (Fig.~\ref{fig:fig5}b), and strong (Fig.~\ref{fig:fig5}c) disorder. Similarly, in Figs.~5d -- 5f we plot  $|S_{12}^N|$ for an array with $N=400$ cells, again for weak (Fig~5d), medium (Fig.~\ref{fig:fig5}e), and strong (Fig.~\ref{fig:fig5}f) disorder. Comparing, for example, Figs.~5a and 5d, corresponding to weak disorder for $N=100$ and $N=400$ cells, respectively, we see that the effect of increasing the size of the array, keeping the same amount of disorder, is to further decrease the transmission, consistent with the localization theory. However, this decrease occurs only away from the teflon resonance frequencies. For medium  and strong disorder, transmission has decayed bellow $10^{-3}$ for most of the frequencies except around the teflon resonances. Thus the localization is not homogeneous at all: the teflon Fabry-Perot resonances strongly suppress the localization.

\section{Localization Length}

In this section we derive an analytical expression for the \emph{localization length} $L_{loc}$, and relate it with the experimental data.
In the case of weak positional disorder,
\begin{equation}\label{small-var}
(k_a\sigma)^2\ll1,
\end{equation}
an analytical expression for this quantity can be obtained as follows.
First, we expand the transfer matrix $Q(n)$ defined by Eq.~\eqref{Q-def},
up to quadratic terms in the perturbation parameter $k_a\sigma\eta(n)$,
\begin{equation}\label{Q-expan}
\hat{Q}(n)\approx
\left\{1-\frac{(k_a\sigma)^2\eta^2(n)}{2}\right\}
\hat{Q}^{(0)}+k_a\sigma\eta(n)\hat{Q}^{(1)}.
\end{equation}
The unperturbed $\hat{Q}^{(0)}$ and first-order $\hat{Q}^{(1)}$ matrices are suitable to be presented in the form
\begin{subequations}\label{Q01-def}
\begin{equation}\label{Q01}
\hat{Q}^{(0)}=
\left(\begin{array}{cc}u&v^*\\[6pt]
v&u^*\end{array}\right),\qquad
\hat{Q}^{(1)}=
\left(\begin{array}{cc}iu&-iv^*\\[6pt]
iv&-iu^*\end{array}\right);
\end{equation}
\begin{eqnarray}
u=[\cos(k_bd_{b})+i\alpha_+\sin(k_bd_{b})]\exp(ik_ad_{a}),
\label{u-def}\\[6pt]
v=i\alpha_-\sin(k_bd_{b})\exp(ik_ad_{a}),\label{v-def}\\[6pt]
\det\hat{Q}^{(0)}=\det\hat{Q}^{(1)}=|u|^2-|v|^2=1.\label{det1}
\end{eqnarray}
\end{subequations}
Also, it is useful for further calculations to take into account that the real and imaginary parts, $u_r\equiv\mathrm{Re}u$ and $u_i\equiv\mathrm{Im}u$, of the matrix element $u$ can be expressed as
\begin{equation}\label{UrUi}
u_r=\cos(\kappa d),\qquad
u_i^2=\sin^2(\kappa d)+|v|^2.
\end{equation}
The first equality is identical to the dispersion relation \eqref{DispRel}, while the second one is a direct consequence of the matrix unimodularity \eqref{det1}.

In order to extract the effects that are solely due to disorder, it is conventional to perform a canonical transformation to the Bloch normal-mode representation in the transfer relation \eqref{An+1An-exp},
\begin{subequations}\label{AA-Blm}
\begin{eqnarray}
\left(\begin{array}{c}\widetilde{A}^{+}_{n+1}\\
\widetilde{A}^{-}_{n+1}\end{array}\right)=
\hat{P}\hat{Q}\hat{P}^{-1}\left(\begin{array}{c}\widetilde{A}^{+}_n\\
\widetilde{A}^{-}_n\end{array}\right),\label{Apm-Blm}\\[6pt]
\left(\begin{array}{c}\widetilde{A}^{+}_n\\
\widetilde{A}^{-}_n\end{array}\right)=
\hat{P}\left(\begin{array}{c}A^{+}_n\\A^{-}_n\end{array}\right).
\label{Atilde-A}
\end{eqnarray}
\end{subequations}
The transformation matrix $\hat{P}$ is specified in such a manner  to make the unperturbed matrix $\hat{Q}^{(0)}$ diagonal,
\begin{equation}\label{Q0-diag}
\hat{P}\hat{Q}^{(0)}\hat{P}^{-1}=
\left(\begin{array}{cc}\exp(+i\kappa d)&0\\[6pt]
0&\exp(-i\kappa d)\end{array}\right),
\end{equation}
in complete accordance with the Floquet
theorem~\cite{flo1883}, or the same, the Bloch
condition~\cite{blo29}. The solution of the problem for the eigenvectors and eigenvalues of $\hat{Q}^{(0)}$ results in
\begin{subequations}\label{P-def}
\begin{eqnarray}
&&\hat{P}=\left(\begin{array}{cc}|v|/\beta_{+}&-iv^*/\beta_{-}\\[6pt]
iv/\beta_{-}&|v|/\beta_{+}\end{array}\right),
\label{P}\\[6pt]
&&\beta_{\pm}^2=2\sqrt{1-u_r^2}\,(u_i\mp\sqrt{1-u_r^2})\nonumber\\[6pt]
&&=2\sin(\kappa d)[u_i\mp\sin(\kappa d)],\label{beta-def}\\[6pt]
&&\beta_{+}^2\beta_{-}^2=4|v|^2\sin^2(\kappa d), \label{beta-beta}\\[6pt]
&&\det\hat{P}=\det\hat{P}^{-1}= |v|^2(\beta_{+}^{-2}-\beta_{-}^{-2})=1.
\label{DetP}
\end{eqnarray}
\end{subequations}

After substituting Eqs.~\eqref{Q-expan}, \eqref{Q01} and \eqref{P} into the canonical transfer relation \eqref{Apm-Blm}, one can obtain the explicit perturbative recursion relations for the new complex amplitudes,
\begin{eqnarray}
\widetilde{A}^{+}_{n+1}&=&\left[1-\frac{k_a^2\sigma^2\eta^2(n)}{2}
+\frac{ik_a\sigma\eta(n)u_i}{\sin(\kappa d)}\right]
\exp(i\kappa d)\widetilde{A}^{+}_{n}
\nonumber\\[6pt]
&&-\frac{k_a\sigma\eta(n)v^*}{\sin(\kappa d)}
\exp(i\kappa d)\widetilde{A}^{-}_{n},
\label{rda-A+}\\[6pt]
\widetilde{A}^{-}_{n+1}&=&\left[1-\frac{k_a^2\sigma^2\eta^2(n)}{2}
-\frac{ik_a\sigma\eta(n)u_i}{\sin(\kappa d)}\right]
\exp(-i\kappa d)\widetilde{A}^{-}_{n}
\nonumber\\[6pt]
&&-\frac{k_a\sigma\eta(n)v}{\sin(\kappa d)}
\exp(-i\kappa d)\widetilde{A}^{+}_{n}.
\label{rda-A-}
\end{eqnarray}

Now one can see from these equations that one equation can be directly obtained from the other just by complex conjugation, if we suppose that $\widetilde{A}^{+}_{n}=\widetilde{A}^{-*}_{n}$. In other words, it is convenient to seek the amplitudes $\widetilde{A}^{\pm}_{n}$ in terms of action-angle
variables,
\begin{equation}\label{A-RTheta}
\widetilde{A}^{\pm}_{n}=R_n\exp(\pm i\theta_n).
\end{equation}
In order to derive the equation for the real amplitude $R_n$, we multiply Eq.~(\ref{rda-A+}) by Eq.~(\ref{rda-A-}). Within the second order of approximation in the perturbation parameter $k_a\sigma\eta(n)$, we have
\begin{eqnarray}\label{rda-Rn}
&&\frac{R^2_{n+1}}{R^2_n}=1+
\frac{2k_a\sigma\eta(n)|v|}{\sin(\kappa d)}\sin(2\theta_n+k_ad_a)
-k^2_a\sigma^2\eta^2(n)
\nonumber\\[6pt]
&&+\frac{k^2_a\sigma^2\eta^2(n)}{\sin^2(\kappa d)} [u_i^2+|v|^2+2u_i|v|\cos(2\theta_n+k_ad_a)].
\end{eqnarray}
The logarithm of Eq.~(\ref{rda-Rn}), that determines the localization length, is also be expanded within the quadratic approximation,
\begin{eqnarray}\label{rda-lnRn}
\ln\left(\frac{R^2_{n+1}}{R^2_n}\right)=
\frac{2k_a\sigma\eta(n)|v|}{\sin(\kappa d)}\sin(2\theta_n+k_ad_a)
\nonumber\\[6pt]
+\frac{2k^2_a\sigma^2\eta^2(n)|v|^2}{\sin^2(\kappa d)}
\Big[1-\sin^2(2\theta_n+k_ad_a)
\nonumber\\[6pt]
+\frac{u_i}{|v|}\cos(2\theta_n+k_ad_a)\Big].
\end{eqnarray}

Now we are in a position to write down the expression for
the inverse localization length $L_{loc}^{-1}$ that is known
to be defined as follows \cite{{izr98}},
\begin{equation}\label{Lyap-def}
L_{loc}^{-1}=
\frac{1}{2d}\langle\overline{\ln\left(\frac{R^2_{n+1}}{R^2_n}\right)}\rangle.
\end{equation}
The average $\langle ab\rangle$ is performed over the disorder
$\eta(n)$ and the average $\overline{ab}$ is carried out over the rapid random phase $\theta_n$. Within the accepted approximation and for \emph{uncorrelated} disorder, see Eq.~\eqref{corr-sym}, we may regard the random quantities $\eta(n)$ and $\eta^2(n)$ to be uncorrelated with trigonometrical functions, containing the angle variable $\theta_n$. Moreover, it can be shown (see, e.g., Ref.~\cite{izr98}) that the distribution of phase $\theta_n$, within the first order of approximation in a weak disorder, is homogeneous (the corresponding distribution function is constant). Therefore, after averaging over $\theta_n$ of Eq.~\eqref{rda-lnRn}, the term linear in $\eta(n)$  and the last term in the brackets vanish and $\sin^2(2\theta_n+k_ad_a)$ is replaced with $1/2$. As a result,
we get
\begin{equation}\label{Lloc-fin}
L_{loc}^{-1}=(k_a\sigma)^2 \frac{\alpha_{-}^2\sin^2(k_bd_b)}{2d\sin^2(\kappa d)}
\end{equation}
This expression is in complete correspondence with that obtained in Refs.~\cite{bal85} and \cite{izr09} using a different approach, and reduces to Eq.~(13) of Ref.~\cite{izr01a} for the limiting case of delta-like barriers.
The appearance of the term  $\sin^2(k_bd_b)$ in the numerator of
Eq.~(\ref{Lloc-fin}) indicates that at frequencies obeying the teflon resonance condition $k_bd_b=m\pi$, the localization length turns into infinity. That is, the random array becomes transparent and this is what is observed in the experimental and numerical
transmission coefficient plotted in Figs.~3-5.

As is known, the localization length is directly related to the transmittance $T_N$ for a finite array of the length $L=Nd$, according to the famous relation $\langle\ln T_N\rangle=-2L/L_{loc}$. In view of this relation and recalling that  $T_N=|S_{12}^N|^2$, see Eq.~\eqref{TN-gen}, it is convenient to introduce the \emph{rescaled} inverse localization length $l_{loc}^{-1}$ as
\begin{equation}\label{avelog-S12}
\langle\ln|S_{12}^N|\rangle=-L/L_{loc}\equiv-l_{loc}^{-1}.
\end{equation}
According to Eq.~\eqref{Lloc-fin}, one can get
\begin{equation}\label{avelog-F}
l_{loc}^{-1}=(\sigma/d)^2F(\nu)N.
\end{equation}
Here we introduced the form-factor
\begin{equation}\label{F-def}
F(\nu)=(k_ad)^2\,
\frac{\alpha_-^2\sin^2(k_bd_b)}{2\sin^2(\kappa d)}
\end{equation}
that specifies the frequency profile of the inverse localization length and is determined only by the parameters of the underlying regular array. The rescaled inverse localization length,\eqref{avelog-F}, increases linearly with the number of cells $N$
and quadratically with the amount of disorder $\sigma/d$.

\begin{figure}
\includegraphics[width=\columnwidth,clip]{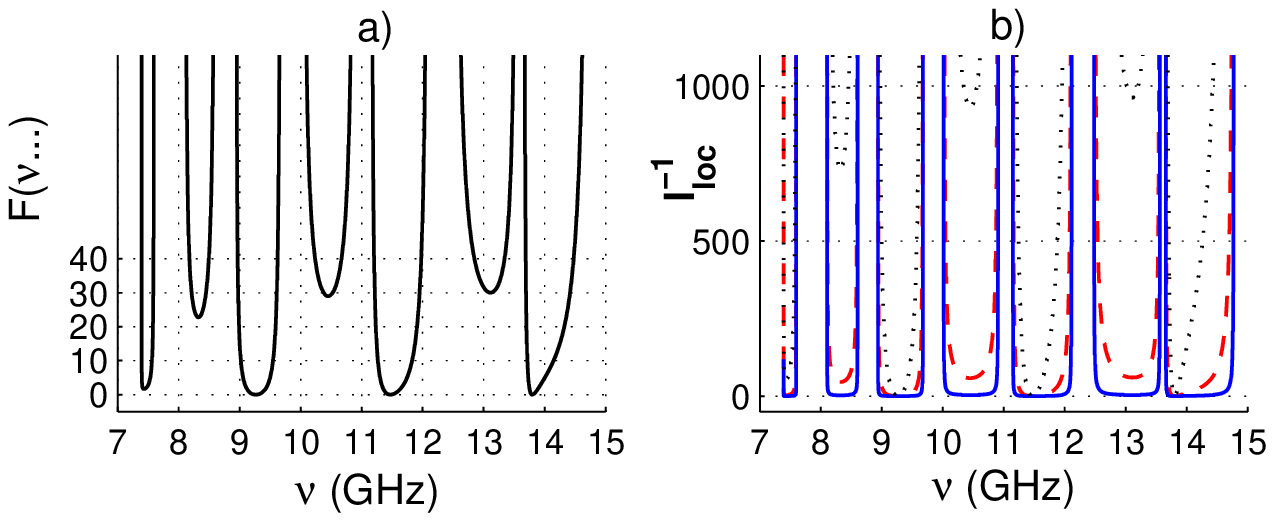}
\includegraphics[width=\columnwidth,clip]{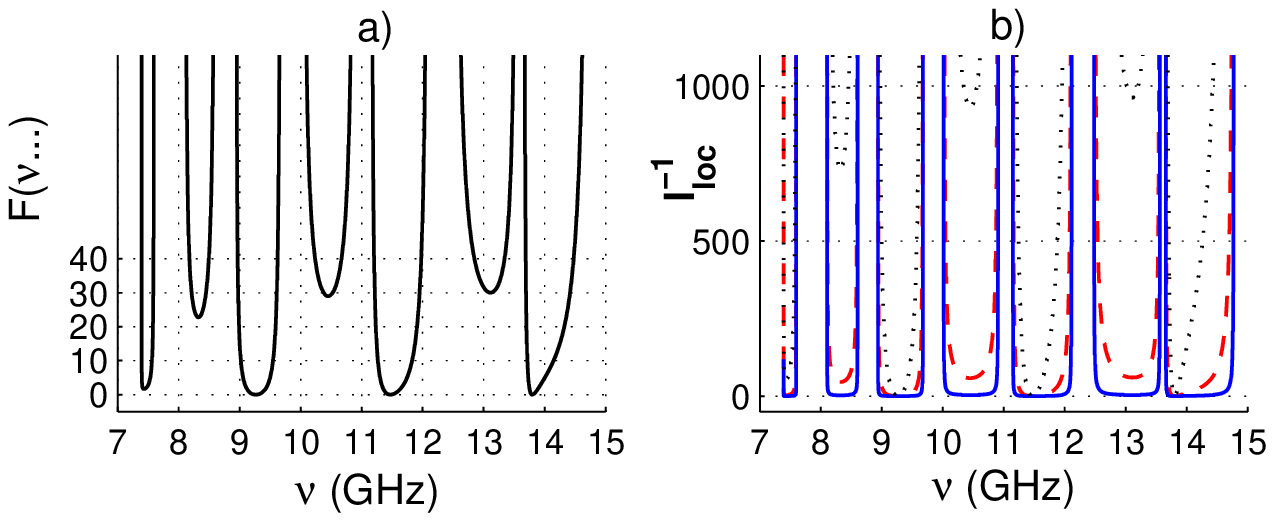}
\caption{\label{fig:fig6}(color online) (a) Form-factor $F(\nu, d_a, d_b, n_a, n_b)$ for
$d_a=d_b=4.078$ cm, $n_a=1$ and $n_b=\sqrt{2.08}$; (b) inverse localization length $l^{-1}_{loc}$ for $N=400$ for weak (solid curve), medium (dashed curve), and strong (dotted curve) disorder.}
\end{figure}

In Fig.~\ref{fig:fig6}a, we plot the form factor $F(\nu)$ for the parameters of our system, $d_a=d_b=4.078$ cm, $n_a=1$, $n_b=\sqrt{2.08}$. In Fig.~\ref{fig:fig6}b, we present the frequency dependence of $l_{loc}^{-1}$ for an array of 400 cells for the cases of weak, medium, and strong disorder. These figures show that the attenuation of transmission is larger for the 6th band and least for the first band, and that the states are completely extended at the teflon resonances.

In order to compare our analytical results with the experimental and numerical ones, we define the theoretical value
$|S_{12}^N|_{th}$ of the transmission spectrum as follows,
\begin{equation}\label{ave-S12}
|S_{12}^N|_{th} \equiv \exp(-l^{-1}_{loc})= \exp[-(\sigma/d)^2F(\nu)N].
\end{equation}
This is what is plotted in green dashed lines in  Fig.~\ref{fig:fig4} for the cases of weak, medium and strong disorder in the $N=26$ cells array and in Fig.~\ref{fig:fig5} for the array with $N=100$ and $N=400$.

Inspection of Figs.~4-5 reveals that the theoretical expression for the transmission spectrum provides a very good description for the case of weak disorder for the whole range of frequencies. For medium disorder, the agrement is moderate but only up to the region of the first
resonance band. The higher the frequency the higher the discrepancy, in accordance with the range of validity \eqref{small-var} of the expression \eqref{avelog-F}, namely, $(k_a\sigma)^2 \ll 1$.

We would like to stress that the \emph{statistically averaged} analytical results of this section required the structure to be sufficiently long. Therefore, a better correspondence with numerics and experiment is expected for longer samples. The data in Figs.~\ref{fig:fig4} and \ref{fig:fig5} confirm this fact.
Indeed, in contrast with $N=26$ (Fig.~\ref{fig:fig4}), for $N=100$ and $N=400$ (Fig.~\ref{fig:fig5}) we can see much better agreement between theory and experiments/numerics for longer chains. The quantity $\exp(-l_{loc}^{-1})$  gives a good fit to the numerical data even for the case of medium disorder and up to $\nu=11.5$ GHz. Observe also that already for $N=100$ the non-resonance bands have completely disappeared for the medium disorder case.

\section{Summary}

We studied the effects of the uncorrelated positional disorder in the 1D Kronig-Penney model paying main attention to the transmission
for weak, medium, and strong disorder. We compare experimental data obtained in the microwave Kronig-Penney setup with the direct calculations based on the transfer matrix method. We show that in spite of a very strong absorbtion in the metallic walls of waveguide, the experimental frequency dependence of the transmission is similar to that obtained numerically by neglecting the absorbtion effect. In particulary, the position of the $N-1$ Fabry-Perot resonances, as well as the teflon resonances are in correspondence with numerical data. The structured similarity between experimental and numerical data can be explained by the different nature of the absorbtion in comparison with the resonance effects. Namely, the latter are due coherent effects, in contrast with the non-coherent nature of the absorbtion.

Another part of our study is related to the expression for the localization length obtained in the perturbative approach for an infinite chain of scattering barriers. Although in the experiment we have a relatively small number of 26 teflon barriers, we show that the expression for the localization length can be effectively used for the description of the transmission coefficient through the finite sample of barriers. To do this, we have used the relation between the transmission coefficient and localization length that involves the finite size of samples, and applied it to the experimental situation. As a result, we have found that the analytical expression for the transmission coefficient reproduces quite well the frequency dependence of experimentally obtained data. Note again, that the global correspondence between analytical and experimental results occurs in the presence of strong absorbtion that is non-avoidable experimentally. The observed correspondence between experimental and analytical results indicate that the expression for the localization length is a working quantity for finite samples, a fact that is not commonly used in applications.

The method we used to derive the expression for the localization length can be generalized to a more general case of the correlated disorder \cite{izr99,izr01b,izr04a,izr05a,her08,izr01a,kuh00a,kro02,izr09}, for which anomalous effects in the transmission are predicted and experimentally observed. Our results can be applied to photonic crystals and electron superlattices, as well as to propagation of acoustic waves in disordered systems.

G.A.L-A acknowledges support from CONACyT, convenio P51458, F.M.I acknowledges financial support from CONACyT grant F2030, and U.~K. and H.J.~S. from the Deutsche Forschungsgemeinschaft via the Forschergruppe 760 ``Scattering Systems with Complex Dynamics''.


\begin{thebibliography}{26}
\expandafter\ifx\csname natexlab\endcsname\relax\def\natexlab#1{#1}\fi
\expandafter\ifx\csname bibnamefont\endcsname\relax
  \def\bibnamefont#1{#1}\fi
\expandafter\ifx\csname bibfnamefont\endcsname\relax
  \def\bibfnamefont#1{#1}\fi
\expandafter\ifx\csname citenamefont\endcsname\relax
  \def\citenamefont#1{#1}\fi
\expandafter\ifx\csname url\endcsname\relax
  \def\url#1{\texttt{#1}}\fi
\expandafter\ifx\csname urlprefix\endcsname\relax\def\urlprefix{URL }\fi
\providecommand{\bibinfo}[2]{#2}
\providecommand{\eprint}[2][]{\url{#2}}

\bibitem[{\citenamefont{Marko\v{s} and Soukoulis}(2008)}]{mar08}
\bibinfo{editor}{\bibfnamefont{P.}~\bibnamefont{Marko\v{s}}} \bibnamefont{and}
  \bibinfo{editor}{\bibfnamefont{C.~M.} \bibnamefont{Soukoulis}}, eds.,
  \emph{\bibinfo{title}{Wave Propagation: {F}rom Electrons to Photonic Crystals
  and Left-Handed Materials}} (\bibinfo{publisher}{Princeton University Press},
  \bibinfo{address}{Princeton}, \bibinfo{year}{2008}).

\bibitem[{\citenamefont{McGurn et~al.}(1993)\citenamefont{McGurn, Christensen,
  Mueller, and Maradudin}}]{mcg93}
\bibinfo{author}{\bibfnamefont{A.~R.} \bibnamefont{McGurn}},
  \bibinfo{author}{\bibfnamefont{K.~T.} \bibnamefont{Christensen}},
  \bibinfo{author}{\bibfnamefont{F.~M.} \bibnamefont{Mueller}},
  \bibnamefont{and} \bibinfo{author}{\bibfnamefont{A.~A.}
  \bibnamefont{Maradudin}}, \bibinfo{journal}{Phys. Rev. B}
  \textbf{\bibinfo{volume}{47}}, \bibinfo{pages}{13120} (\bibinfo{year}{1993}).

\bibitem[{\citenamefont{Smith et~al.}(2000)\citenamefont{Smith, Padilla, Vier,
  Nemat-Nasser, and Schultz}}]{smi00}
\bibinfo{author}{\bibfnamefont{D.~R.} \bibnamefont{Smith}},
  \bibinfo{author}{\bibfnamefont{W.~J.} \bibnamefont{Padilla}},
  \bibinfo{author}{\bibfnamefont{D.~C.} \bibnamefont{Vier}},
  \bibinfo{author}{\bibfnamefont{S.~C.} \bibnamefont{Nemat-Nasser}},
  \bibnamefont{and} \bibinfo{author}{\bibfnamefont{S.}~\bibnamefont{Schultz}},
  \bibinfo{journal}{Phys. Rev. Lett.} \textbf{\bibinfo{volume}{84}},
  \bibinfo{pages}{4184} (\bibinfo{year}{2000}).

\bibitem[{\citenamefont{Shelby et~al.}(2001)\citenamefont{Shelby, Smith, and
  Schultz}}]{she01}
\bibinfo{author}{\bibfnamefont{R.~A.} \bibnamefont{Shelby}},
  \bibinfo{author}{\bibfnamefont{D.~R.} \bibnamefont{Smith}}, \bibnamefont{and}
  \bibinfo{author}{\bibfnamefont{S.}~\bibnamefont{Schultz}},
  \bibinfo{journal}{Science} \textbf{\bibinfo{volume}{292}},
  \bibinfo{pages}{77} (\bibinfo{year}{2001}).

\bibitem[{\citenamefont{Parazzoli et~al.}(2003)\citenamefont{Parazzoli,
  Greegor, Li, Koltenbah, and Tanielian}}]{par03}
\bibinfo{author}{\bibfnamefont{C.~G.} \bibnamefont{Parazzoli}},
  \bibinfo{author}{\bibfnamefont{R.~B.} \bibnamefont{Greegor}},
  \bibinfo{author}{\bibfnamefont{K.}~\bibnamefont{Li}},
  \bibinfo{author}{\bibfnamefont{B.~E.~C.} \bibnamefont{Koltenbah}},
  \bibnamefont{and}
  \bibinfo{author}{\bibfnamefont{M.}~\bibnamefont{Tanielian}},
  \bibinfo{journal}{Phys. Rev. Lett.} \textbf{\bibinfo{volume}{90}},
  \bibinfo{pages}{107401} (\bibinfo{year}{2003}).

\bibitem[{\citenamefont{Esmailpour et~al.}(2006)\citenamefont{Esmailpour,
  Esmaeilzadeh, Faizabadi, Carpena, and Tabar}}]{esm06}
\bibinfo{author}{\bibfnamefont{A.}~\bibnamefont{Esmailpour}},
  \bibinfo{author}{\bibfnamefont{M.}~\bibnamefont{Esmaeilzadeh}},
  \bibinfo{author}{\bibfnamefont{E.}~\bibnamefont{Faizabadi}},
  \bibinfo{author}{\bibfnamefont{P.}~\bibnamefont{Carpena}}, \bibnamefont{and}
  \bibinfo{author}{\bibfnamefont{M.~R.~R.} \bibnamefont{Tabar}},
  \bibinfo{journal}{Phys. Rev. B} \textbf{\bibinfo{volume}{74}},
  \bibinfo{pages}{024206} (\bibinfo{year}{2006}).

\bibitem[{\citenamefont{Nau et~al.}(2007)\citenamefont{Nau, Sch\"{o}nhardt,
  Bauer, Christ, Zentgraf, Kuhl, Klein, and Giessen}}]{nau07}
\bibinfo{author}{\bibfnamefont{D.}~\bibnamefont{Nau}},
  \bibinfo{author}{\bibfnamefont{A.}~\bibnamefont{Sch\"{o}nhardt}},
  \bibinfo{author}{\bibfnamefont{C.}~\bibnamefont{Bauer}},
  \bibinfo{author}{\bibfnamefont{A.}~\bibnamefont{Christ}},
  \bibinfo{author}{\bibfnamefont{T.}~\bibnamefont{Zentgraf}},
  \bibinfo{author}{\bibfnamefont{J.}~\bibnamefont{Kuhl}},
  \bibinfo{author}{\bibfnamefont{M.~W.} \bibnamefont{Klein}}, \bibnamefont{and}
  \bibinfo{author}{\bibfnamefont{H.}~\bibnamefont{Giessen}},
  \bibinfo{journal}{Phys. Rev. Lett.} \textbf{\bibinfo{volume}{98}},
  \bibinfo{pages}{133902} (\bibinfo{year}{2007}).

\bibitem[{\citenamefont{Ponomarev et~al.}(2007)\citenamefont{Ponomarev, Schwab,
  Dasbach, Bayer, Reinecke, Reithmaier, and Forchel}}]{pon07}
\bibinfo{author}{\bibfnamefont{I.~V.} \bibnamefont{Ponomarev}},
  \bibinfo{author}{\bibfnamefont{M.}~\bibnamefont{Schwab}},
  \bibinfo{author}{\bibfnamefont{G.}~\bibnamefont{Dasbach}},
  \bibinfo{author}{\bibfnamefont{M.}~\bibnamefont{Bayer}},
  \bibinfo{author}{\bibfnamefont{T.~L.} \bibnamefont{Reinecke}},
  \bibinfo{author}{\bibfnamefont{J.~P.} \bibnamefont{Reithmaier}},
  \bibnamefont{and} \bibinfo{author}{\bibfnamefont{A.}~\bibnamefont{Forchel}},
  \bibinfo{journal}{Phys. Rev. B} \textbf{\bibinfo{volume}{75}},
  \bibinfo{pages}{205434} (\bibinfo{year}{2007}).

\bibitem[{\citenamefont{Asatryan et~al.}(2007)\citenamefont{Asatryan, Botten,
  Byrne, Freilikher, Gredeskul, Shadrivov, McPhedran, and Kivshar}}]{asa07}
\bibinfo{author}{\bibfnamefont{A.~A.} \bibnamefont{Asatryan}},
  \bibinfo{author}{\bibfnamefont{L.~C.} \bibnamefont{Botten}},
  \bibinfo{author}{\bibfnamefont{M.~A.} \bibnamefont{Byrne}},
  \bibinfo{author}{\bibfnamefont{V.~D.} \bibnamefont{Freilikher}},
  \bibinfo{author}{\bibfnamefont{S.~A.} \bibnamefont{Gredeskul}},
  \bibinfo{author}{\bibfnamefont{I.~V.} \bibnamefont{Shadrivov}},
  \bibinfo{author}{\bibfnamefont{R.~C.} \bibnamefont{McPhedran}},
  \bibnamefont{and} \bibinfo{author}{\bibfnamefont{Y.~S.}
  \bibnamefont{Kivshar}}, \bibinfo{journal}{Phys. Rev. Lett.}
  \textbf{\bibinfo{volume}{99}}, \bibinfo{pages}{193902}
  (\bibinfo{year}{2007}).

\bibitem[{\citenamefont{Dong and Xiong}(1998)}]{don98b}
\bibinfo{author}{\bibfnamefont{H.}~\bibnamefont{Dong}} \bibnamefont{and}
  \bibinfo{author}{\bibfnamefont{S.-J.} \bibnamefont{Xiong}},
  \bibinfo{journal}{J. Phys.: Condens. Matter} \textbf{\bibinfo{volume}{10}},
  \bibinfo{pages}{7691} (\bibinfo{year}{1998}).

\bibitem[{\citenamefont{Baluni and Willemsen}(1985)}]{bal85}
\bibinfo{author}{\bibfnamefont{V.}~\bibnamefont{Baluni}} \bibnamefont{and}
  \bibinfo{author}{\bibfnamefont{J.}~\bibnamefont{Willemsen}},
  \bibinfo{journal}{Phys. Rev. A} \textbf{\bibinfo{volume}{31}},
  \bibinfo{pages}{3358} (\bibinfo{year}{1985}).

\bibitem[{\citenamefont{Luna-Acosta et~al.}(2008)\citenamefont{Luna-Acosta,
  Schanze, Kuhl, and St{\"o}ckmann}}]{lun08}
\bibinfo{author}{\bibfnamefont{G.~A.} \bibnamefont{Luna-Acosta}},
  \bibinfo{author}{\bibfnamefont{H.}~\bibnamefont{Schanze}},
  \bibinfo{author}{\bibfnamefont{U.}~\bibnamefont{Kuhl}}, \bibnamefont{and}
  \bibinfo{author}{\bibfnamefont{H.-J.} \bibnamefont{St{\"o}ckmann}},
  \bibinfo{journal}{New J. of Physics} \textbf{\bibinfo{volume}{10}},
  \bibinfo{pages}{043005} (\bibinfo{year}{2008}).

\bibitem[{\citenamefont{Kuhl and St{\"o}ckmann}(1998)}]{kuh98b}
\bibinfo{author}{\bibfnamefont{U.}~\bibnamefont{Kuhl}} \bibnamefont{and}
  \bibinfo{author}{\bibfnamefont{H.-J.} \bibnamefont{St{\"o}ckmann}},
  \bibinfo{journal}{Phys. Rev. Lett.} \textbf{\bibinfo{volume}{80}},
  \bibinfo{pages}{3232} (\bibinfo{year}{1998}).

\bibitem[{\citenamefont{Kuhl et~al.}(2000)\citenamefont{Kuhl, Izrailev,
  Krokhin, and St{\"o}ckmann}}]{kuh00a}
\bibinfo{author}{\bibfnamefont{U.}~\bibnamefont{Kuhl}},
  \bibinfo{author}{\bibfnamefont{F.~M.} \bibnamefont{Izrailev}},
  \bibinfo{author}{\bibfnamefont{A.~A.} \bibnamefont{Krokhin}},
  \bibnamefont{and} \bibinfo{author}{\bibfnamefont{H.-J.}
  \bibnamefont{St{\"o}ckmann}}, \bibinfo{journal}{Appl. Phys. Lett.}
  \textbf{\bibinfo{volume}{77}}, \bibinfo{pages}{633} (\bibinfo{year}{2000}).

\bibitem[{\citenamefont{Krokhin et~al.}(2002)\citenamefont{Krokhin, Izrailev,
  Kuhl, St{\"o}ckmann, and Ulloa}}]{kro02}
\bibinfo{author}{\bibfnamefont{A.}~\bibnamefont{Krokhin}},
  \bibinfo{author}{\bibfnamefont{F.}~\bibnamefont{Izrailev}},
  \bibinfo{author}{\bibfnamefont{U.}~\bibnamefont{Kuhl}},
  \bibinfo{author}{\bibfnamefont{H.-J.} \bibnamefont{St{\"o}ckmann}},
  \bibnamefont{and} \bibinfo{author}{\bibfnamefont{S.~E.} \bibnamefont{Ulloa}},
  \bibinfo{journal}{Physica E} \textbf{\bibinfo{volume}{13}},
  \bibinfo{pages}{695} (\bibinfo{year}{2002}), \bibinfo{note}{conf. on
  Modulated Semicond. Structures (MSS 10)}.

\bibitem[{\citenamefont{Kuhl et~al.}(2008)\citenamefont{Kuhl, Izrailev, and
  Krokhin}}]{kuh08a}
\bibinfo{author}{\bibfnamefont{U.}~\bibnamefont{Kuhl}},
  \bibinfo{author}{\bibfnamefont{F.~M.} \bibnamefont{Izrailev}},
  \bibnamefont{and} \bibinfo{author}{\bibfnamefont{A.~A.}
  \bibnamefont{Krokhin}}, \bibinfo{journal}{Phys. Rev. Lett.}
  \textbf{\bibinfo{volume}{100}}, \bibinfo{pages}{126402}
  (\bibinfo{year}{2008}).

\bibitem[{\citenamefont{Floquet}(1883)}]{flo1883}
\bibinfo{author}{\bibfnamefont{G.}~\bibnamefont{Floquet}},
  \bibinfo{journal}{Ann. de l'Ecole Normale} \textbf{\bibinfo{volume}{12}},
  \bibinfo{pages}{47} (\bibinfo{year}{1883}).

\bibitem[{\citenamefont{Bloch}(1929)}]{blo29}
\bibinfo{author}{\bibfnamefont{F.}~\bibnamefont{Bloch}}, \bibinfo{journal}{Z.
  Phys. A} \textbf{\bibinfo{volume}{52}}, \bibinfo{pages}{555}
  (\bibinfo{year}{1929}).

\bibitem[{\citenamefont{Izrailev et~al.}(1998)\citenamefont{Izrailev, Ruffo,
  and Tessieri}}]{izr98}
\bibinfo{author}{\bibfnamefont{F.~M.} \bibnamefont{Izrailev}},
  \bibinfo{author}{\bibfnamefont{S.}~\bibnamefont{Ruffo}}, \bibnamefont{and}
  \bibinfo{author}{\bibfnamefont{L.}~\bibnamefont{Tessieri}},
  \bibinfo{journal}{J. Phys. A} \textbf{\bibinfo{volume}{31}},
  \bibinfo{pages}{5263} (\bibinfo{year}{1998}).

\bibitem[{\citenamefont{Izrailev and Makarov}(2009)}]{izr09}
\bibinfo{author}{\bibfnamefont{F.~M.} \bibnamefont{Izrailev}} \bibnamefont{and}
  \bibinfo{author}{\bibfnamefont{N.~M.} \bibnamefont{Makarov}},
  \bibinfo{journal}{Phys. Rev. Lett.} \textbf{\bibinfo{volume}{102}},
  \bibinfo{pages}{203901} (\bibinfo{year}{2009}).

\bibitem[{\citenamefont{Izrailev et~al.}(2001)\citenamefont{Izrailev, Krokhin,
  and Ulloa}}]{izr01a}
\bibinfo{author}{\bibfnamefont{F.~M.} \bibnamefont{Izrailev}},
  \bibinfo{author}{\bibfnamefont{A.~A.} \bibnamefont{Krokhin}},
  \bibnamefont{and} \bibinfo{author}{\bibfnamefont{S.~E.} \bibnamefont{Ulloa}},
  \bibinfo{journal}{Phys. Rev. B} \textbf{\bibinfo{volume}{63}},
  \bibinfo{pages}{041102} (\bibinfo{year}{2001}).

\bibitem[{\citenamefont{Izrailev and Krokhin}(1999)}]{izr99}
\bibinfo{author}{\bibfnamefont{F.~M.} \bibnamefont{Izrailev}} \bibnamefont{and}
  \bibinfo{author}{\bibfnamefont{A.~A.} \bibnamefont{Krokhin}},
  \bibinfo{journal}{Phys. Rev. Lett.} \textbf{\bibinfo{volume}{82}},
  \bibinfo{pages}{4062} (\bibinfo{year}{1999}).

\bibitem[{\citenamefont{Izrailev and Makarov}(2001)}]{izr01b}
\bibinfo{author}{\bibfnamefont{F.~M.} \bibnamefont{Izrailev}} \bibnamefont{and}
  \bibinfo{author}{\bibfnamefont{N.~M.} \bibnamefont{Makarov}},
  \bibinfo{journal}{Opt. Lett.} \textbf{\bibinfo{volume}{26}},
  \bibinfo{pages}{1604} (\bibinfo{year}{2001}).

\bibitem[{\citenamefont{Izrailev and Makarov}(2004)}]{izr04a}
\bibinfo{author}{\bibfnamefont{F.~M.} \bibnamefont{Izrailev}} \bibnamefont{and}
  \bibinfo{author}{\bibfnamefont{N.~M.} \bibnamefont{Makarov}},
  \bibinfo{journal}{Appl. Phys. Lett.} \textbf{\bibinfo{volume}{84}},
  \bibinfo{pages}{5150} (\bibinfo{year}{2004}).

\bibitem[{\citenamefont{Izrailev and Makarov}(2005)}]{izr05a}
\bibinfo{author}{\bibfnamefont{F.~M.} \bibnamefont{Izrailev}} \bibnamefont{and}
  \bibinfo{author}{\bibfnamefont{N.~M.} \bibnamefont{Makarov}},
  \bibinfo{journal}{J. Phys. A} \textbf{\bibinfo{volume}{38}},
  \bibinfo{pages}{10613} (\bibinfo{year}{2005}).

\bibitem[{\citenamefont{Herrej{\'o}n et~al.}(2008)\citenamefont{Herrej{\'o}n,
  Izrailev, and Tessieri}}]{her08}
\bibinfo{author}{\bibfnamefont{J.~C.~H.} \bibnamefont{Herrej{\'o}n}},
  \bibinfo{author}{\bibfnamefont{F.~M.} \bibnamefont{Izrailev}},
  \bibnamefont{and} \bibinfo{author}{\bibfnamefont{L.}~\bibnamefont{Tessieri}},
  \bibinfo{journal}{Physica E} \textbf{\bibinfo{volume}{40}},
  \bibinfo{pages}{3137} (\bibinfo{year}{2008}).

\end{thebibliography}
\end{document}